\documentclass[9pt, letterpaper]{article}
\usepackage{amsmath}
\usepackage{graphicx}
\usepackage{xurl}
\usepackage[left=2cm, right=2cm, top=2cm, bottom=2cm]{geometry}
\usepackage{hyperref}
\usepackage{booktabs}
\usepackage{longtable}
\usepackage{rotating}
\usepackage{cite}

\hypersetup{
    colorlinks=true,
    linkcolor=blue,
    citecolor=blue,
    urlcolor=blue,
    pdftitle={Signing Right Away (SRA)},
    pdfpagemode=FullScreen,
}

\title{Signing Right Away}
\author{Yejun Jang}
\date{\today}

\begin{document}

\maketitle

\begin{abstract}
The proliferation of high-fidelity synthetic media, coupled with exploitable hardware vulnerabilities in conventional imaging pipelines, has precipitated a crisis of trust in digital content. Existing countermeasures, from post-hoc classifiers to software-based signing, fail to address the fundamental challenge of establishing an unbreakable link to reality at the moment of capture. This whitepaper introduces Signing Right Away (SRA), a comprehensive security architecture that guarantees the provenance of digital media from ``silicon to silicon to signed file.'' SRA leverages a four-pillar security model—Confidentiality, Integrity, Authentication, and Replay Protection, akin to the MIPI Camera Security Framework (CSF), but also extends its scope beyond the internal data bus to the creation of a cryptographically sealed, C2PA-compliant final asset. By securing the entire imaging pipeline within a Trusted Execution Environment (TEE), SRA ensures that every captured image and video carries an immutable, verifiable proof of origin. This provides a foundational solution for industries reliant on trustworthy visual information, including journalism, legal evidence, and insurance. We present the SRA architecture, a detailed implementation roadmap informed by empirical prototyping, and a comparative analysis that positions SRA as the essential ``last mile'' in the chain of content trust.
\end{abstract}

\section{Motivation}

The digital information ecosystem is at a critical inflection point. The confluence of two powerful trends—the democratization of hyper-realistic synthetic media generation and the persistent neglect of fundamental hardware security in imaging devices—has created an environment where the authenticity of visual content can no longer be taken for granted. This section establishes the urgent need for a new paradigm in content authentication by examining the dual nature of this threat and the inadequacy of current solutions.

\subsection{The Erosion of Trust}

The advent of sophisticated generative models has fundamentally altered the landscape of digital content creation. Technologies such as OpenAI's DALL-E series for images \cite{ramesh2021zeroshot, ramesh2022hierarchical, betker2023improving}, Sora for video \cite{videoworldsimulators2024}, and Google's AudioLM for audio \cite{borsos2023audiolm} have made the generation of high-fidelity synthetic content accessible to a global user base with minimal technical expertise. What once required specialized skills and significant computational resources can now be accomplished with simple text prompts.

While these tools offer unprecedented creative potential, they also represent a potent new vector for disinformation and fraud. The ability to generate convincing but entirely fabricated visual ``evidence'' poses a direct threat to social cohesion, financial markets, and democratic processes. A stark illustration of this threat occurred in May 2023, when an AI-generated image depicting an explosion near the Pentagon was circulated on social media by a verified account\cite{OSullivanPassantino2023}. The viral spread of this fabricated image triggered a brief but significant dip in the S\&P 500 index, demonstrating the tangible, real-world consequences of synthetic media weaponized for informational warfare. This incident is not an isolated anomaly but a precursor of a new reality where the foundational assumption of ``seeing is believing'' is irrevocably broken. The core problem is the erosion of a shared visual reality, undermining the very basis of trust in digital communication.

\subsection{The Hardware Attack Surface}

Compounding the threat from generative AI is a critical, yet widely overlooked, vulnerability at the hardware level of nearly all modern imaging devices, from smartphones to professional cameras. The connection between the image sensor and the host system's Image Signal Processor (ISP) is typically managed by the MIPI Camera Serial Interface 2 (CSI-2) protocol \cite{mipi_csi2}. In its standard implementation, the raw bitstream of pixel data transmitted over this interface is unencrypted and unauthenticated.

This architectural flaw creates a gaping security hole. It allows for a straightforward man-in-the-middle attack, where a malicious hardware component can intercept the data path and inject a completely fabricated video feed, which the host system will accept as genuine data from its own camera sensor. This is not a theoretical vulnerability; it is a practical one, enabled by commercially available hardware. Devices marketed as ``HDMI to CSI-2 adapters,'' often based on bridge chips like the Toshiba TC358743XBG, are sold openly and can be used to perform precisely this type of data injection attack \cite{hdmi_csi_adapter, toshiba_hdmi_csi_bridge}. An attacker can connect any HDMI source—such as a computer playing a synthetically generated video—to one of these adapters, which then outputs a valid CSI-2 signal that can be fed directly into a device's motherboard, bypassing the camera entirely.

The synergy between these two threats creates a perfect storm for sophisticated deception. An adversary can first use a generative model to create a hyper-realistic fake video of a significant event. Next, they can use an off-the-shelf HDMI-to-CSI-2 adapter to inject this video feed into a legitimate, trusted device (e.g., a specific model of smartphone). The device's operating system, having no reason to distrust the data arriving at its ISP, would process the video as if it were captured by its own sensor. If the device is running a content authentication application, it could then proceed to generate a cryptographically valid signature and a C2PA manifest for the fake content. The final result is a piece of synthetic media bearing a legitimate digital signature from a trusted hardware source. This combined attack vector circumvents nearly all conventional methods of verification, demonstrating that any solution that does not address security at the hardware source is fundamentally incomplete.

\section{Related Works}

In response to the ``fake data'' problem, several countermeasures have been proposed and deployed. However, a systematic analysis reveals that each suffers from limitations that render them insufficient to counter the combined hardware and software threat vector described above.

\subsection{Neural Network Classifiers}
One approach involves training deep learning models to discriminate between real and synthetic content. While these classifiers can achieve some success under controlled conditions, they are inherently brittle. Their performance degrades significantly when faced with common transformations like compression, which is ubiquitous on the internet \cite{rossler2019faceforensics}. Furthermore, they struggle to generalize to content produced by novel generator architectures they were not trained on \cite{chai2020what}. The very nature of Generative Adversarial Networks (GANs), which co-evolve a generator and a discriminator (a classifier), means that generative models are actively trained to defeat such detection methods, locking classifiers in a perpetual and likely unwinnable arms race.

\subsection{Digital Watermarking}
Digital watermarking involves embedding a hidden, noise-like signal into media to track its origin or identify it as synthetic. Technologies like Google's SynthID are a prominent example of this approach, designed to mark AI-generated content in a way that is imperceptible to humans but detectable by a specific algorithm \cite{google_synthid}. While useful for copyright enforcement and providing a ``label'' for synthetic media, watermarking is not a robust security mechanism. Watermarks can be fragile and may be damaged or removed by subsequent image processing, compression, or even simple cropping. More importantly, watermarking is a post-hoc labeling technique; it does nothing to prevent the initial injection of unsigned, malicious data at the hardware level.

\subsection{Software-Level Digital Signing (Standard C2PA)}
The Coalition for Content Provenance and Authenticity (C2PA) has established a vital open standard for creating interoperable, cryptographically signed metadata that records the provenance of a digital asset \cite{c2pa_spec_1_3}. A C2PA manifest can detail an asset's creation, authorship, and editing history, providing a transparent record for consumers \cite{c2pa_explainer}. However, when implemented purely in software on an untrusted device, the C2PA standard has a critical limitation: it can only attest to the provenance of the digital data it is given. A software-based C2PA tool signs a manifest that says, ``This application, running on this device, received these pixels and is now signing them.'' It has no way of verifying that those pixels originated from the device's actual camera sensor. This creates a ``garbage in, gospel out'' scenario. If a fake video stream is injected via a hardware attack, a software C2PA implementation will dutifully sign it, lending a false aura of authenticity to the fabricated content. The signature is cryptographically valid, but the underlying asset is a lie.

As summarized in Table \ref{tab:solutions-comparison}, existing solutions fail to provide a complete, resilient defense because they do not operate at the true point of origin: the hardware interface between the sensor and the processor. This is the critical gap that the Signing Right Away architecture is designed to fill.

\begin{table}
\centering
\footnotesize
\label{tab:solutions-comparison}
\begin{tabular}{@{}p{3.0cm} p{3.0cm} p{2.5cm} p{3.5cm} p{3.5cm}@{}}
\toprule
\textbf{Solution} & \textbf{Mechanism} & \textbf{Point of Application} & \textbf{Resilience to Tampering} & \textbf{Primary Use Case} \\
\midrule
Neural Classifier & Statistical pattern recognition to detect artifacts of generation. & Post-Processing / Distribution & Low. Fails on novel generators and compressed content. & Detecting known types of synthetic media. \\
\addlinespace
Digital Watermarking & Embedding a hidden signal (e.g., SynthID) into the media file. & Post-Processing / At Generation & Medium. Can be degraded by transformations and is not a security feature. & Labeling and tracking content (e.g., copyright). \\
\addlinespace
Software C2PA & Cryptographic signature over a manifest of metadata. & Post-Capture / Software Layer & High (for the manifest), but Zero for the content's origin. & Interoperable provenance logging in untrusted environments. \\
\addlinespace
\textbf{SRA} & \textbf{Hardware-secured capture pipeline with authenticated encryption and TEE-based signing.} & \textbf{Point of Capture (Sensor-to-SoC)} & \textbf{Very High. Protects against hardware injection and software compromise.} & \textbf{Establishing verifiable, hardware-rooted proof of origin.} \\
\bottomrule
\end{tabular}
\caption{Comparative Analysis of Content Authenticity Solutions}
\end{table}

\section{Signing Right Away (SRA)}

To address the fundamental vulnerabilities outlined in the previous section, a new architectural approach is required. Signing Right Away (SRA) is a comprehensive security framework designed to establish an unbreakable chain of trust that extends from the moment photons strike the image sensor to the final, portable, and independently verifiable digital asset. It is not merely a signing protocol but an end-to-end secure imaging pipeline that integrates hardware-level protection with standardized, cryptographically-verifiable provenance records.

\subsection{Core Principles}

The foundational philosophy of SRA is to create an unforgeable, hardware-enforced cryptographic link between a physical event and its digital representation ``right away'' at the moment of capture. The architecture is built on the premise that trust must be established at the earliest possible point in the imaging pipeline and maintained through every subsequent stage of processing. By securing the data path from ``silicon to silicon"—from the image sensor's silicon to the host System-on-Chip's (SoC) silicon—SRA eliminates the hardware attack surface that plagues conventional systems. This hardware-rooted trust is then extended to the final file, creating a complete ``photon to file'' chain of custody.

\subsection{A Four-Pillar Security Model}

The SRA architecture is structured around four essential security pillars. The design was originally conceived based on fundamental cryptographic principles and authenticated encryption schemes like ChaCha20-Poly1305. During development, we discovered that the MIPI Camera Security Framework had independently standardized similar approaches, validating our architectural choices. Each pillar addresses a specific threat to the integrity of the imaging pipeline.

\begin{itemize}
    \item \textbf{Authentication:} The process begins with a mutual cryptographic handshake between the image sensor module and the host SoC. This step ensures that only genuine, authorized hardware components are part of the pipeline. It cryptographically verifies the identity of the camera, preventing the connection of rogue sensors or malicious data injection hardware, such as the HDMI-to-CSI-2 adapters previously discussed. This establishes the trusted endpoints of the secure channel.
    
    \item \textbf{Confidentiality:} Once trust is established, all image data transmitted over the MIPI CSI-2 bus is encrypted using a strong, authenticated stream cipher (e.g., AES-GCM). This renders the raw camera feed unintelligible to any unauthorized party attempting to eavesdrop on the physical connection, protecting the privacy and content of the data while it is in transit.
    
    \item \textbf{Integrity:} Alongside encryption, every packet or frame of image data is accompanied by a Message Authentication Code (MAC). This cryptographic tag is calculated by the sensor and verified by the SoC. Any modification to the data in transit, no matter how small, will cause the MAC verification to fail. This guarantees that the data received by the SoC is bit-for-bit identical to the data that left the sensor, preventing any form of in-transit tampering or pixel manipulation.
    
    \item \textbf{Replay Protection:} To defend against an attacker capturing a valid, encrypted frame and re-injecting it into the stream at a later time, the protocol incorporates anti-replay mechanisms. This is typically achieved through the use of monotonically increasing sequence numbers or unique nonces for each frame, which are included in the data used to calculate the MAC. The receiving SoC will reject any frame that is out of sequence or has a repeated nonce, ensuring the ``liveness'' of the data stream.
\end{itemize}

\subsection{Trusted Execution Environment (TEE)}

While the four pillars secure the data in transit, the processing and signing of the data must occur in a protected environment on the host SoC. SRA designates a Trusted Execution Environment (TEE)—such as Arm TrustZone or Qualcomm's Secure Execution Environment (QSEE/QTEE)—as the ``secure chamber'' for all sensitive operations. The TEE is a hardware-isolated enclave that runs in parallel to the main operating system (the ``normal world") and has exclusive access to protected hardware and memory.

The TEE's responsibilities within the SRA architecture are critical:
\begin{enumerate}
    \item \textbf{Secure Key Management:} The TEE securely stores and manages all cryptographic keys, including the session keys for the encrypted CSI-2 channel and, most importantly, the device's unique private signing key. This private key is generated, stored, and used exclusively within the TEE and is never exposed to the normal world OS or any applications running on it. This design directly mitigates the class of vulnerabilities exploited by security firms like ElcomSoft, who were able to extract private signing keys from the firmware of Canon and Nikon cameras, thereby compromising their entire image authentication systems \cite{elcomsoft_canon, elcomsoft_nikon, cardinalpeak_elcomsoft}.
    
    \item \textbf{Secure Data Processing:} The encrypted and authenticated image stream is passed directly from the ISP to the TEE for decryption and validation. Any necessary image processing, such as format conversion (e.g., RAW to JPEG), occurs within the TEE's isolated memory. This prevents a compromised OS or malicious application from accessing or tampering with the decrypted, unsigned image data.
    
    \item \textbf{Cryptographic Signing:} Once the image is processed, the TEE generates a C2PA manifest containing provenance metadata and signs it using the device's private key. The TEE then binds this signed manifest to the final image file before releasing it to the normal world for storage or transmission.
\end{enumerate}
By performing these operations within the TEE, SRA ensures that an unsigned, authentic image never exists in an environment accessible to potentially malicious software. The only asset that ever leaves the secure chamber is the final, cryptographically sealed product.

\subsection{Integration with C2PA}

For the provenance record to be useful, it must be universally accessible and understandable. SRA leverages the C2PA open standard as its final output format, ensuring that any image signed by an SRA-enabled device is immediately interoperable with a global ecosystem of C2PA-compliant viewers and verification tools \cite{c2pa_spec_main}.

The true power of SRA lies in the unique assertions it embeds within the C2PA manifest. A standard software-based C2PA manifest might record that an image was created by ``Device X.'' An SRA-generated manifest provides a much stronger, verifiable claim. It contains cryptographic proof not just of \textit{what} created the image, but \textit{how} it was created. The manifest generated by the SRA TEE would include specific assertions such as:
\begin{itemize}
    \item \textbf{Hard Binding:} A cryptographic hash (e.g., SHA-256) of the final image pixels, ensuring that any subsequent alteration to the image will be detected \cite{c2pa_spec_main}.
    \item \textbf{Device Identity and Attestation:} A statement signed with the device's unique private key, which can be linked to a certificate from the manufacturer, attesting to the device's authenticity. The manifest can also include a TEE attestation, proving that the signing operation was performed by specific, certified firmware running within the hardware-secure enclave.
    \item \textbf{Secure Pipeline Metadata:} Custom C2PA assertions that explicitly state the security context of the capture. For example: ``This image was captured via a secure pipeline,'' with details like ``Encryption: AES-GCM-128,'' ``Authentication: Succeeded,'' and a unique session or frame counter.
    \item \textbf{Secure Ancillary Data:} Timestamps and geolocation data sourced from trusted hardware components and passed directly to the TEE, rather than being sourced from the potentially compromised normal world OS.
\end{itemize}
This transforms the C2PA manifest from a simple, self-asserted log file into a verifiable security audit trail. A third-party verifier does not need to trust the device's software; they can cryptographically verify the proof embedded in the manifest. They can see evidence that the image originated from an authenticated sensor, traveled through a protected hardware path, was processed in a secure enclave, and was signed by a key that is cryptographically bound to the device manufacturer. This provides a level of assurance that is an order of magnitude stronger than what software-only solutions can offer.

\section{Contribution}

To fully appreciate the contribution of the SRA architecture, it is essential to compare it with the MIPI Alliance's Camera Security Framework (CSF), a landmark specification that provides the foundational elements for securing in-device camera pipelines. While SRA was independently developed and later converged to similar architectural principles as the MIPI CSF, it serves a distinct and complementary purpose, extending the chain of trust beyond the device to the end consumer of the media.

\subsection{MIPI Camera Security Framework (CSF) Overview}

The MIPI Camera Security Framework was developed primarily to address the stringent security and safety requirements of the automotive industry, particularly for Advanced Driver-Assistance Systems (ADAS) and autonomous driving applications \cite{mipi_csf_whitepaper, mipi_csf_press_release}. In these safety-critical systems, it is paramount that the vehicle's electronic control unit (ECU) can trust that the data it receives from cameras and other sensors is authentic and has not been tampered with \cite{mipi_csf_main}.

The framework consists of a suite of specifications designed to provide end-to-end, application-level security for data transmitted over MIPI CSI-2 \cite{mipi_csf_main, mipi_csf_press_release}. Its key components include:
\begin{itemize}
    \item \textbf{MIPI Camera Service Extensions (CSE) v2.0:} This specification defines the protocols for embedding security information directly into the CSI-2 data stream. It offers two main methods: Service Extensions Packet (SEP), which adds security headers and footers to each data packet, and Frame-Based Service Extensions Data (FSED), which appends dedicated security packets to an image frame. These protocols carry the Message Authentication Codes (MACs) necessary for integrity protection \cite{mipi_csf_main}.
    \item \textbf{MIPI Camera Security v1.0:} This specification defines the system-level management of security, leveraging the industry-standard DMTF Security Protocol and Data Model (SPDM) to enable mutual authentication of system components and establish secure sessions \cite{mipi_csf_main}.
    \item \textbf{MIPI Camera Security Profiles v1.0:} To ensure interoperability, this document defines specific cryptographic suites. The two primary profiles are the ``efficiency'' ciphersuite, which provides integrity-only protection using AES-CMAC, and the ``performance'' ciphersuite, which provides both integrity (AES-GMAC) and optional confidentiality (AES-CTR encryption) for systems with dedicated hardware support \cite{mipi_csf_main, synopsys_mipi_cse}.
\end{itemize}
The MIPI CSF is a comprehensive and flexible solution for its intended purpose: guaranteeing that the data sink (e.g., an SoC or ECU) within a closed system receives authentic data from the data source (the image sensor).

\subsection{Distiction and Project Scope}

The fundamental distinction between the MIPI CSF and SRA lies in their scope and ultimate objective. They address different parts of the overall trust problem.

The \textbf{MIPI Camera Security Framework's scope is internal}, focused on securing the data pipeline \textit{within} a device or system. Its primary goal is to protect the \textit{process} of data transmission from the sensor to the processor. The trust boundary of the MIPI CSF effectively ends at the SoC. It ensures the SoC can trust its inputs, but it makes no provisions for how a third party can trust the SoC's outputs.

The \textbf{Signing Right Away architecture's scope is external}, focused on creating a portable, self-contained, and independently verifiable digital asset. Its primary goal is to protect the final \textit{product}—the image or video file—and imbue it with a verifiable provenance that can be trusted by any third party, anywhere in the world, long after it has left the original device. SRA extends the trust boundary from the SoC to the global information ecosystem.

An effective analogy is to consider the process of minting currency. The MIPI CSF is akin to the armored truck and secure protocols used to transport raw gold (the image data) from the mine (the sensor) to the national mint (the SoC). Its purpose is to prevent theft (eavesdropping) or substitution (tampering) along the way. SRA, by contrast, is the process that happens \textit{inside} the mint. It takes the authenticated gold, verifies its purity, and then stamps it into a serialized, certified coin (the signed C2PA file) whose authenticity can be recognized and trusted by anyone. SRA is the essential ``last mile'' that transforms a securely transmitted internal data stream into a globally trusted asset.

\subsection{Securing Intellectual Property}

This distinction in scope makes SRA a powerful tool for securing intellectual property and addressing copyright challenges in ways that a pipeline security protocol alone cannot. By creating a signed, verifiable asset at the point of capture, SRA provides creators with robust, cryptographically-enforceable evidence of authorship and originality.

\begin{itemize}
    \item \textbf{Unforgeable Proof of Authorship:} The digital signature on the C2PA manifest is generated using the device's unique, TEE-protected private key. This signature serves as unforgeable proof that a specific device created the content at a specific time. This can be linked to a registered owner or organization, establishing a clear chain of title from the moment of creation.
    
    \item \textbf{Secure Timestamping:} The use of a trusted clock, managed within the TEE, to generate the timestamp in the C2PA manifest provides strong, tamper-evident proof of the creation date. This is a critical piece of evidence in establishing priority in copyright disputes.
    
    \item \textbf{Tamper-Evident Licensing Information:} The C2PA manifest can include assertions for copyright notices, creator information, and machine-readable rights expressions (e.g., Creative Commons licenses). Because this information is part of the cryptographically sealed manifest, it cannot be stripped or altered without invalidating the signature. This ensures that the creator's intended usage rights travel with the content.
\end{itemize}

This architecture provides a potent defense against a sophisticated form of infringement that could be termed ``provenance laundering.'' In a typical scenario, an infringer might take a copyrighted image, apply minor edits, strip its EXIF metadata, and re-publish it as their own original work. With SRA, this becomes significantly more difficult. The original image would possess an immutable C2PA manifest linking it to the true creator and creation time. The infringer's edited version would either have a broken signature (if they kept the manifest but changed the pixels) or no signature at all (if they stripped the manifest). In a future ecosystem where SRA-signed content is common, an image claiming to be original but lacking a valid provenance record would be inherently suspect. This shifts the burden of proof, making it harder for infringing content to be passed off as legitimate and easier for creators to defend their work.

For maximum resilience, SRA can be used in a hybrid approach with digital watermarking. SRA provides the primary, authoritative, and cryptographically secure proof of origin. A robust digital watermark can be added as a secondary, resilient layer for tracking content that has been screenshotted or transcoded—actions that would break the C2PA signature but might leave a watermark intact. The SRA manifest could even include a signed assertion stating, ``A `Type-X' digital watermark was embedded at the time of capture,'' linking the two technologies together.

\section{Development}

The SRA architecture is not merely a theoretical construct; it is the result of a pragmatic development process informed by empirical prototyping and a keen awareness of the evolving technology landscape. The project's evolution from an initial hardware-centric prototype to a mature, ecosystem-aligned framework demonstrates a data-driven approach to solving the complex challenges of content provenance.

\subsection{Initial Prototyping and Empirical Findings}

The initial phase of the SRA project focused on building a proof-of-concept prototype to validate the core concept of real-time cryptographic signing of a camera feed. The fundamental architecture—comprising an authenticated encrypted camera link and immediate signing in a TEE—was established in the original 2024 design based on cryptographic first principles. This prototype consisted of a Sony IMX219 camera module, an Efinix Trion T20 FPGA for processing, and a Raspberry Pi as the receiving host, using the `raspiraw` utility on an older version of its operating system. This phase, rather than being a simple implementation, became a crucial research endeavor that yielded invaluable insights about implementation strategy while the core architectural design remained consistent.

\begin{itemize}
    \item \textbf{Finding 1: The Practical Limits of CSI-2 Metadata.} The initial plan was to transmit the AES-CMAC authentication tag from the FPGA to the Raspberry Pi using the standard `User-Defined 8-bit' data types (0x30-0x37) within the CSI-2 protocol. However, investigation revealed that the Raspberry Pi's CSI-2 receiver (Unicam) did not support these standard types, instead routing any non-image packets to a generic metadata buffer without clear parsing rules. To overcome this, a clever workaround was devised: the CMAC tag was encapsulated in a packet with a valid but unused image data type identifier (e.g., RGB888 - 0x24) and prefixed with a unique custom header byte (0x0B). This allowed the `raspiraw` utility's metadata decoding function to be modified to specifically look for this header and correctly parse the CMAC tag. This experience highlighted the real-world challenges of interoperability and the need for a robust, well-defined protocol for security metadata, as later specified in the MIPI CSE standard.
    
    \item \textbf{Finding 2: The Prohibitive Cost of Real-Time Software Cryptography.} The original design specified authenticated encryption schemes like ChaCha20-Poly1305, but the prototype initially tested a custom AES-CMAC module designed in Verilog for the FPGA. While functional, simulation and testing revealed a significant performance bottleneck. The processing of a single 1920x1232 resolution frame to generate a CMAC tag required approximately ten million clock cycles. To meet the target frame rate of 30 frames per second (fps), a simple calculation revealed the demanding hardware requirement:
    $$ 10,000,000 \frac{\text{cycles}}{\text{frame}} \times 30 \frac{\text{frames}}{\text{second}} = 300,000,000 \frac{\text{cycles}}{\text{second}} $$
    This meant the FPGA's clock frequency would need to be at least 300 MHz, and likely higher to account for overhead, exceeding the practical limits of the designed module on the target hardware. This finding was a decisive moment, proving that real-time, high-resolution authenticated encryption is not feasible with soft-core cryptographic implementations on general-purpose FPGAs. It established a clear requirement for dedicated, hardware-accelerated cryptographic engines to make the SRA vision practical.
\end{itemize}

\subsection{Implementation Roadmap}

The lessons learned from the initial prototype informed the implementation strategy while preserving the core architectural vision. The key insight was that rather than developing custom hardware for every function, SRA should be implemented as firmware on top of existing, commercially available trusted hardware. This realization led to a refined, five-stage development plan designed to systematically build and deploy the full SRA architecture with hardware-accelerated cryptography. This roadmap, summarized in Table \ref{tab:roadmap}, demonstrates a logical progression from baseline functionality to full ecosystem integration.

\begin{sidewaystable}[htbp]
\centering
\label{tab:roadmap}
\begin{tabular}{@{}p{3.5cm} p{2.0cm} p{4.0cm} p{4.0cm} p{4.0cm}@{}}
\toprule
\textbf{Stage} & \textbf{Key Objective} & \textbf{Core Technologies} & \textbf{Validation Milestone} & \textbf{Connection to Prototype Findings} \\
\midrule
1: Baseline & Establish a functional, non-secure camera pipeline. & MIPI CSI-2, FPGA/ISP, Basic image processing (Bayer to RGB, JPEG). & A clear video stream is successfully transmitted from sensor to host. & Re-establishes the baseline system from the prototype with production-oriented design. Part of the original architectural plan. \\
\addlinespace
2: Secure Channel & Implement an authenticated and encrypted camera feed. & Authenticated Encryption with Associated Data (AEAD) cipher (e.g., AES-GCM), mutual authentication protocol (e.g., SPDM). & The host successfully receives, decrypts, and validates the integrity of the encrypted video stream in real time. & Core architectural component from the original design. Prototype findings informed the choice of hardware-accelerated AEAD over soft-core implementations. \\
\addlinespace
3: C2PA Integration & Digitally sign the content and bind provenance metadata. & C2PA Rust SDK, digital signatures (ECDSA/RSA), custom C2PA assertions. & A valid, signed C2PA manifest is generated for each captured image, containing secure pipeline metadata. & Core architectural component from the original design. Implements the immediate signing concept that was always central to SRA. \\
\addlinespace
4: TEE Hardening & Isolate all sensitive operations within a secure enclave. & Arm TrustZone / Qualcomm TEE, secure key storage, TEE attestation. & The entire process from decryption to signing occurs within the TEE, with private keys never exposed to the normal OS. & Provides the secure environment for key management that was outside the scope of the initial hardware-focused prototype. \\
\addlinespace
5: SoC Integration & Port the SRA framework to commercial mobile SoCs. & Qualcomm Secure Camera APIs, Android HAL, OEM partnerships. & SRA functions as a ``Secure Capture Mode'' on a commercial-grade smartphone, producing C2PA-compliant assets. & Represents the strategic shift to leveraging existing trusted hardware, making SRA scalable and commercially viable. \\
\bottomrule
\end{tabular}
\caption{SRA Phased Development Roadmap}
\end{sidewaystable}

\subsection{Aligning with the Broader Ecosystem}

The final stage of the roadmap reflects a critical strategic insight: the most effective path to widespread adoption is not to reinvent the wheel, but to align with and build upon the secure hardware capabilities that are already being integrated into commercial SoCs. While the original five-stage plan included a final phase for designing custom Application-Specific Integrated Circuits (ASICs), the prototype experience and an analysis of market trends revealed that leveraging existing trusted hardware in commercial SoCs was a more efficient path to deployment.

Mobile SoC vendors like Qualcomm have already integrated the necessary hardware primitives—such as secure ISPs, hardware crypto accelerators, and robust TEEs—into their platforms, often driven by the need for features like secure biometric authentication \cite{truepic_foresight_snapdragon}. The emergence of the Qualcomm Snapdragon 8 Gen 3 as the first C2PA-compliant mobile platform is a powerful validation of this trend \cite{truepic_snapdragon}.

Furthermore, industry pioneers like Truepic have already demonstrated a similar architecture in practice. Their Foresight system leverages the Qualcomm TEE and the secure hardware pipeline of the Spectra ISP to create hardware-secured, authenticated photos on Snapdragon-powered devices \cite{truepic_foresight_snapdragon}. This serves as a powerful proof-of-concept for the SRA model and demonstrates a clear path to market.

SRA's strategy is therefore to position itself as an open, interoperable reference architecture that can be implemented on any SoC that provides the necessary trusted hardware components. By leveraging existing secure camera APIs and TEE SDKs, SRA can be deployed as a firmware or software solution that ``lights up'' the latent security capabilities of modern devices. This approach dramatically reduces the cost and time-to-market compared to a custom silicon strategy, and it fosters a competitive, multi-vendor ecosystem rather than a single proprietary solution. This pragmatic pivot, informed by both empirical research and market observation, positions SRA not as a disruptive replacement for existing technologies, but as an essential, value-added layer that is poised for rapid adoption.

\section{Conclusion}

The digital world is contending with an unprecedented crisis of authenticity. The dual threats of hyper-realistic generative AI and fundamental hardware vulnerabilities have systematically dismantled the traditional foundations of trust in visual media. In this new landscape, incremental solutions are insufficient. A new architectural standard is required—one that is rooted in verifiable hardware security and extends trust from the moment of creation to the point of consumption.

The Signing Right Away (SRA) architecture provides this standard. By establishing a cryptographically secure and authenticated channel from the image sensor to a Trusted Execution Environment, SRA closes the critical hardware vulnerability that allows for the injection of fabricated content. By performing all sensitive processing and signing within a hardware-isolated enclave, it protects the integrity of the provenance process from software-level compromise. Finally, by outputting a C2PA-compliant manifest enriched with verifiable security metadata, it creates a portable, interoperable, and globally trusted digital asset.

SRA does not seek to replace existing standards but to complete them. It builds upon the principles of the MIPI Camera Security Framework, extending its internal pipeline protection to create a verifiable external product. It enhances the C2PA standard, transforming its provenance manifests from self-asserted logs into cryptographically-backed security audit trails. It provides a foundational layer of trust upon which other technologies, such as digital watermarking for copyright tracking, can be built.

The development roadmap for SRA, from an ambitious hardware prototype to a mature, ecosystem-aligned framework, demonstrates a commitment to solving this problem in a pragmatic and scalable manner. By aligning with the capabilities of modern mobile SoCs, SRA charts a clear path to widespread adoption. The implementation of this architecture represents a critical step towards restoring a shared sense of visual reality, fostering a future where every critical image and video—whether used in a courtroom, a news report, or an insurance claim—can be fundamentally trusted at its source.

\newpage

\bibliographystyle{plain}
\bibliography{references}

\end{document}